\documentclass[conference,a4paper]{IEEEtran}
\IEEEoverridecommandlockouts
\usepackage{cite}
\usepackage{amsmath,amssymb,amsfonts}
\usepackage{algorithmic}
\usepackage{graphicx}
\usepackage{textcomp}
\usepackage{xcolor}
\usepackage[ruled,linesnumbered]{algorithm2e}
\usepackage{float}
\usepackage{subcaption}
\usepackage{hyperref}
\usepackage{comment}

\def\BibTeX{{\rm B\kern-.05em{\sc i\kern-.025em b}\kern-.08em
    T\kern-.1667em\lower.7ex\hbox{E}\kern-.125emX}}

\begin{document}
\title{Collaborative Inference in DNN-based Satellite Systems with Dynamic Task Streams}

\author{\IEEEauthorblockN{Jinglong Guan$^{1}$, Qiyang Zhang$^{1,2}$, Ilir Murturi$^{2}$, Praveen Kumar Donta$^{2}$, Schahram Dustdar$^{2}$, Shangguang Wang$^{1}$\\
\textit{$^{1}$State Key Laboratory of Networking and Switching Technology},\\
\textit{Beijing University of Posts and Telecommunicatons}, Beijing, China\\
\textit{$^{2}$Distributed System Group, TU Wien}, Vienna, Austria\\
\{mr\_guanjl;qyzhang;sgwang\}@bupt.edu.cn;\{imurturi;pdonta;dustdar\}@dsg.tuwien.ac.at}
}


\maketitle
\begin{abstract}
As a driving force in the advancement of intelligent in-orbit applications, DNN models have been gradually integrated into satellites, producing daily latency-constraint and computation-intensive tasks. However, the substantial computation capability of DNN models, coupled with the instability of the satellite-ground link, pose significant challenges, hindering timely completion of tasks. It becomes necessary to adapt to task stream changes when dealing with tasks requiring latency guarantees, such as dynamic observation tasks on the satellites. To this end, we consider a system model for a collaborative inference system with latency constraints, leveraging the multi-exit and model partition technology. To address this, we propose an algorithm, which is tailored to effectively address the trade-off between task completion and maintaining satisfactory task accuracy by dynamically choosing early-exit and partition points. Simulation evaluations show that our proposed algorithm significantly outperforms baseline algorithms across the task stream with strict latency constraints.

\end{abstract}

\begin{IEEEkeywords}
Satellite inference, Task offloading, Multi-exit DNNs, Model partition
\end{IEEEkeywords}
 
\section{Introduction}

With the continuous technological advancements and increasing drive for space exploration, recent years have witnessed the proliferation of Low Earth Orbit (LEO) satellites such as Telesat and SpaceX \cite{b1}.
Today, around 44\% of in-orbit LEO satellites are dedicated to Earth Observation (EO) purposes \cite{example}.
Many machine learning algorithms, especially Deep Neural Networks (DNN),
are employed for inference raw image observations such as forest fires and geomorphic changes \cite{b18}.

The prevailing approach suggests transmitting all observations to the ground
for image analysis, utilizing a traditional method referred to as the “bent pipe” architecture \cite{denby2020orbital}. However, due to the typically unreliable and intermittent satellite-ground environment, a significant amount of raw observations remains untransmitted in a limited time.
An alternative solution involves executing DNN models directly on the satellite where the data is generated \cite{wang2023satellite}. However, the majority of LEO satellites are equipped with limited computing power and small physical size,
which pose challenges in handling the high computational demands associated with complex DNN models. Much attention has been given to offloading in-orbit computing tasks to the ground as a promising solution. Typically, existing efforts focus on offloading tasks to ground stations for processing, provided that the satellites are within the coverage of specific ground stations. While straightforward and practical, this solution can suffer from a computational bottleneck due to the limited constraints of each satellite. 

To achieve efficient inference offloading, many state-of-the-art systems like Tracking and Data Relay Satellite (TDRS)  and European Data Relay System (EDRS) involve satellite relays in High Earth Orbit (HEO) satellites for task offloading \cite{lai2021orbitcast}. It is reported that even a free-space laser link achieves up to 1.8 Gbps data rate \cite{laux2012status}.
When faced with extensive and diverse computational demands, it is necessary to collaborate the efforts of multiple satellites.
The task owner distributes inference tasks across multiple satellites, with each satellite equipped with high-performance commercial, off-the-shelf (COTS) computational hardware.

Despite these efforts, it remains challenging for the inference task streams within the limited time constraints in potentially extreme factors \cite{lai2021orbitcast} such as poor network conditions and excessive task arrivals.
To meet the requirement of current EO tasks, which can range from demanding precision to partial precision at times, multiple satellites are required to collectively perform observations, with different satellites handling specific computational tasks.
In such a computing environment characterized by diversity and dynamics, multi-exit mechanisms are crucial to address these issues.
Multi-exit mechanisms are introduced as a solution to reduce a significant computational load by selectively activating an early-exit (EE) point \cite{zhang2023comprehensive} within a multi-layer network, even if it leads to inference degradation. Partitioning multi-exit models for multiple task stream applications remains to be fully considered. However, the simultaneous management of EE points and partition points adds complexity to the inference system, especially when dealing with evolving task profiles.

This paper proposes an efficient framework for inference offloading involving multiple satellites by partitioning multi-exit DNN models, which consist of one HEO satellite and multiple LEO satellites.
We divide the inference task into two parts, with the task owner executing the initial portion of the DNNs and offloading the remaining part to other task executors.
The key challenge lies in determining the optimal EE and partitioning points for each task, based on specific task demands and the current system status.
We need to consider the trade-off between task completion and maintaining satisfactory task accuracy. When prioritizing high inference performance for each task, some tasks may fail to meet their deadlines due to prolonged waiting times. In turn, if we solely focus on increasing system throughput, overall inference performance may not meet expectations. Therefore, this paper needs to formulate this issue and design efficient algorithms to address it.

The main contributions of our paper include the following aspects.
\begin{itemize}
\item 
We formulate an optimization problem by considering inference task and computing models to trade-off between task completion and maintaining satisfactory task accuracy under the time limitations.
\item
We introduce a task gain-aware method based on task completion and task accuracy, aiming to derive an efficient solution to the formulated problem.
\item Simulation evaluations show that our proposed algorithm significantly outperforms the baseline algorithms across the task stream.
\end{itemize}

\section{Background and Related Work}
\textbf{“Bent pipe” architecture.} 
Most LEO satellites have recently employed a “bent pipe” architecture, where satellites function as data-transmitting relays rather than data-processing capabilities. For example,
Giuliari et al. \cite{b5} introduced an inter-networking approach that involves ground stations serving as gateways for satellite-ground communications, aimed at enhancing satellite data transmission efficiency.
Cheng et al. \cite{b4} proposed an integrated satellite-ground computing architecture that utilizes satellites to enable access to cloud resources on the ground. 
However, data transmission between satellites and the ground is significantly hindered by an unstable communication link and intermittent availability. Consequently, this results in a substantially lower transmission rate compared to the rate at which data is generated onboard, leading to a noteworthy increase in the overall latency of all tasks.

\textbf{In-orbit computing.}
This paradigm has attracted much attention in the satellite computing domain.
Denby et al. \cite{denby2020orbital} proposed the orbital edge computing framework that performs data filtering on the satellites, which aims to enhance downlinking efficiency in saturated satellites while considering limited computational capacity. 
Denby et al. \cite{denby2023kodan} recently presented a novel computing architecture that maximizes the utility of saturated satellite-ground downlinking while alleviating computational bottlenecks.
Therefore, for existing computing schemes, the entire inference process is executed on satellites. 
The limited computing resources and low energy acquisition capabilities of this approach also place a heavy burden on satellites. 
To address these issues, 
we investigate a model partitioning strategy for fine-grained computational resource optimization based on AI-oriented tasks.

\textbf{Model partitioning for offloading inference.} 
Some efforts suggest that model partitioning can be used to offload inference tasks from task owners \cite{he2020joint,miao2020adaptive}. Neurosurgeon \cite{kang2017neurosurgeon} proposed layer-wise partition strategies to adaptively offload model computation. JointDNN \cite{eshratifar2019jointdnn} was a collaboration framework that employs both model training and inference to achieve resource optimization.
Nonetheless, these works ensure model partitions to guarantee inference accuracy under various constraints.

\textbf{Multi-exit models for boosting task inference.} There has been a growing interest in partitioning multi-exit models to enhance the performance of DNNs.
For example, EDeepSave \cite{ju2021edeepsave} aimed to improve the inference performance by the EE mechanism to avoid interruption when handing over to the mobile network.
Edgent \cite{li2019edge} proposed a model of DNN partitioning to maximize inference accuracy while satisfying latency constraints. Different from the above, we consider multiple inference task streams to balance task latency and accuracy.

\section{System Model and Problem Formulation}\label{sec:modeling}
\begin{figure}[t]
  \centering
  \begin{subfigure}[b]{\linewidth}
    \centering
    \includegraphics[width=\linewidth]{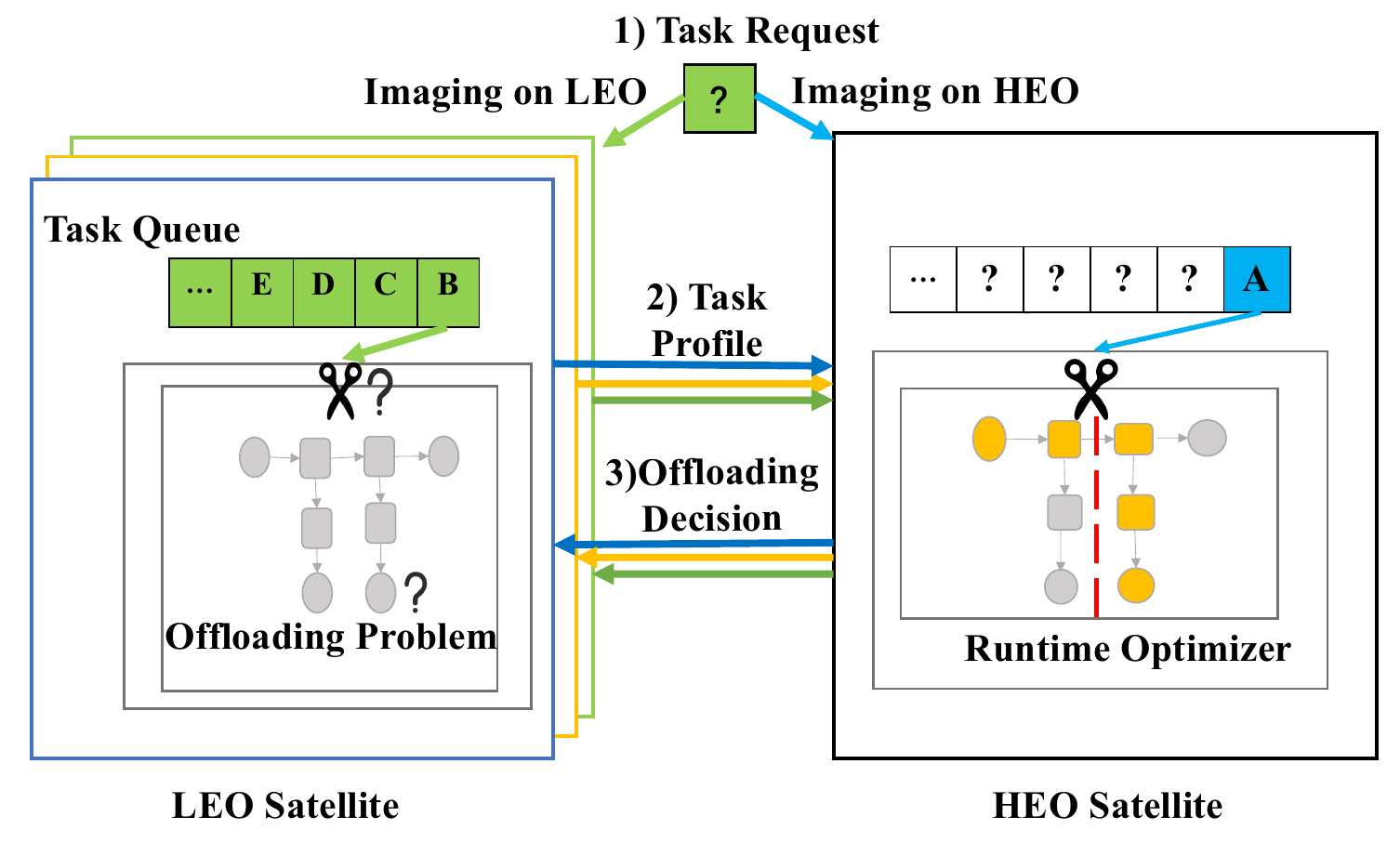}
    \caption{Decision stage.}
    \label{fig:subfig1}
  \end{subfigure}
  \quad
  \begin{subfigure}[b]{\linewidth}
    \centering
    \includegraphics[width=\linewidth]{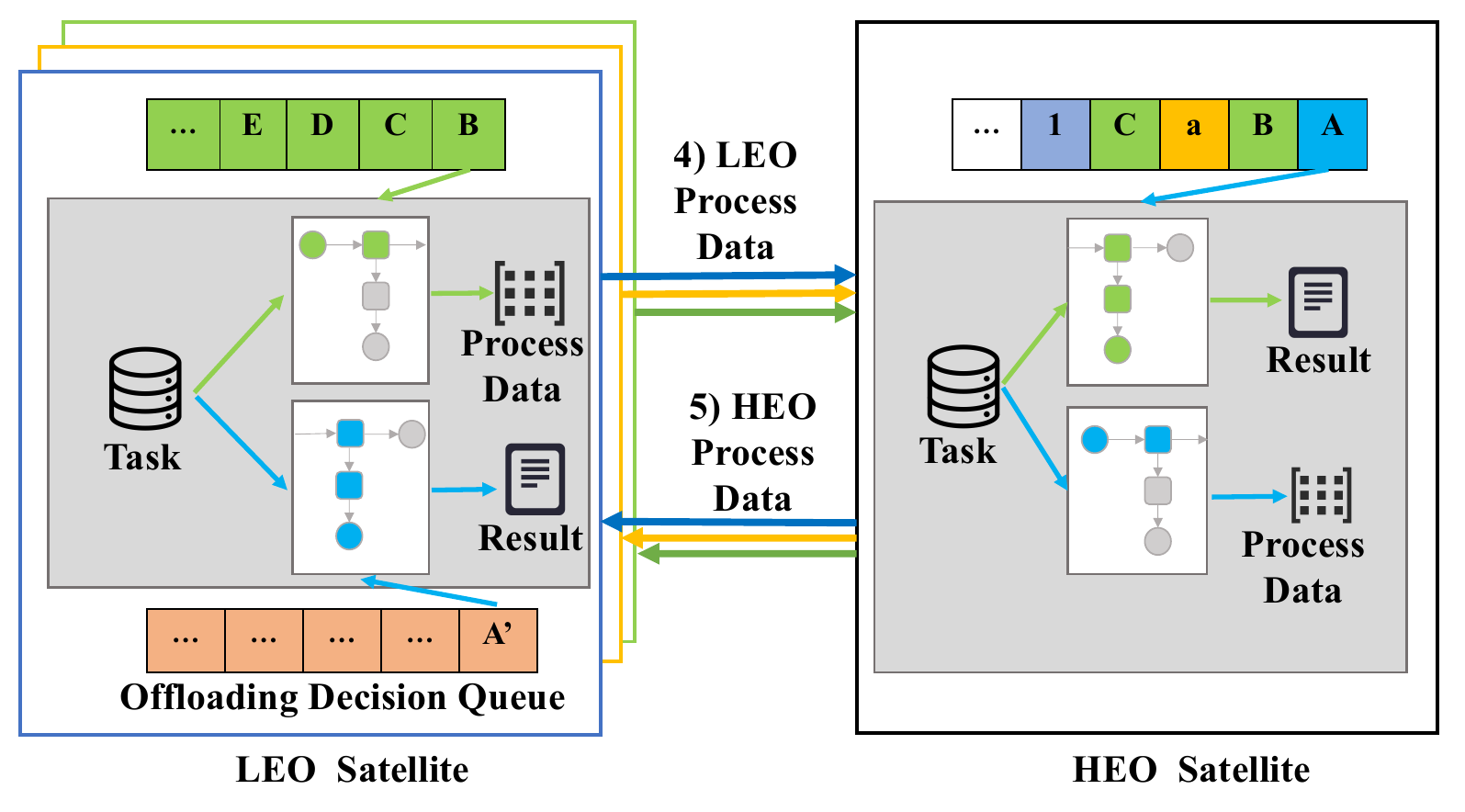}
    \caption{Implementation stage.}
    \label{fig:subfig2}
  \end{subfigure}
  \caption{The workflow of collaborative task stream offloading system.}
  \label{fig:main}
\end{figure}

Fig. \ref{fig:main} illustrates the time-slotted task stream system, which includes one HEO satellite and multiple LEO satellites.
For each satellite, we deployed the pre-trained models for subsequent adaptive inference tasks in advance.
HEO satellite images typically feature high spatial resolution and extensive coverage, while LEO satellite images offer even greater resolution and more frequent observations. The selection of the appropriate satellite depends on the specific task requirements.
We describe the workflow of an inference system involving two main stages as follows: 
1) A task request includes current task requirements when it arrives. Specifically, for HEO imaging tasks, the HEO satellite captures the image first and stores it in the task queue; for LEO imaging tasks, the LEO satellite transmits the task information (e.g., task type, data size, time constraint) to the HEO satellite.
In the case of imaging tasks, the HEO satellite determines offloading solutions and transmits the offloading decision back to the LEO satellites; 
2) When LEO satellites perform imaging tasks, they receive offloading information, process the front part, and send the remaining part to HEO satellites. HEO satellites handle these tasks on the First-Come-First-Served (FCFS) principle.
During HEO imaging tasks, HEO satellites also process the front part and then send the rest to LEO satellites; LEO satellites receive the processing result once HEO satellites complete the task.
For ease of discussion, we do not consider the time it takes for the final results from the HEO satellite for LEO imaging tasks \cite{song2021adaptive}.

\subsection{Inference Task Model}

Let the task be denoted by $T_n$ = ($S_n$, $D_n$, $L_n$, $t^{A}_{n}$), where $T_n$ indicates the $n$-th task information in the task stream, $S_n$ indicates task type, $D_n$ indicates task size, $L_n$ indicates the maximum allowable task time constraint, and $t^A_{n}$ indicates the arrival time. Given that there is a proportional relationship between $L_n$ and $D_n$, we have $L_n = k\times{D_n}, k\in N$. Meanwhile, the number of EE points is $M$ and the layer number in the $i$-th EE point is $M_i$. Further, the EE and the partition points of the $n$-th task are $E_n$ and $P_n$ respectively, with the constraints $0 \leq{E_n}\leq{M}$, ${0}\leq{P_n}\leq{M_{E_n}}$.  

Inspired by \cite{eshratifar2019jointdnn}, processing time and transmission data size during handover are application-specific based on profiling. The prediction model can provide accurate predictions in practical applications, characterized by high reliability and minimal implementation costs.
Consequently, we use a linear regression-based prediction model, which can accurately estimate the inference time and output data size of each layer in the multi-exit model, considering both LEO and HEO satellites. For example, we consider the case of LEO satellites. We utilize \textit{timeline} to record the inference time and fit the inference time of each model layer with a linear function. For the input data size $D_n$, $f^d_{i,j}(D_n)$ represents the predicted inference time of the $j$th layer in the $i$th EE point, which is executed on the LEO satellite if $d = 0$ or the HEO satellite if $d = 1$. $f^d_{i,j}(D_n)$ represents the predicted output data size of the $j$th layer in the $i$th EE point of the model.

\subsection{Computing Model}
When selecting $E_n$ and $P_n$, we divide the entire system into three stages: First, when the imaging task is executed on the LEO, we compute task $T_n$’s first stage processing time as $t^{P_1}_n=\sum_{i=1}^{P_n}{{f^0_{{E_n},i}}(D_n)}$, or $t^{P_1}_n=0$; Second, when imaging task is executed on the LEO, we can compute task $T_n$’s second stage processing time as $t^{P_2}_n=\sum_{i={P_n}+1}^{M_{E_n}}{{f^1_{{E_n},i}}(D_n)}$, or $t^{P_2}_n=\sum_{i=1}^{P_n}{{f^1_{{E_n},i}}(D_n)}$; Third, when imaging task is executed on the HEO, we can compute task $T_n$’s third stage processing time as $t^{P_3}_n=0$, or $t^{P_3}_n=\sum_{i={P_n}+1}^{M_{E_n}}{{f^0_{{E_n},i}}(D_n)}$.

We denote the transmission time of inter-satellite links as $L^{ISL}_{latency}$, which is denoted as follows:
\begin{align}
    L^{ISL}_{latency} = \frac{len\left(D\right)}{w\left({S_{L}},{{S_{H}}}\right)} + \frac{Q_{(S_{L},S_{H})}}{c}
\end{align}
where $S_{L}$ denotes the LEO satellite, $S_{H}$ denotes the HEO satellite, $D$ denotes the transmitted data size between LEO and HEO, $Q_{(S_{L},S_{H})}$ denotes the length of the transmission link between HEO and LEO, $c$ denotes the speed of laser communication. We denote the transmission rate of data $w\left({S_{L}},{{S_{H}}}\right)$ as follows:
\begin{align}
    w\left({S_{L}},{{S_{H}}}\right) = B\log_{2}{\left({1 + {\frac{P_r{\left({S_L},{S_H}\right)}}{{k_B}{T_s}{B}{\gamma}}}}\right)}
\end{align}
where $B$ denotes the bandwidth, $k_B$ denotes the \textit{Boltzmann's constant}, $T_s$ denotes the system noise temperature, $\gamma$ denotes the Signal-to-Noise ratio, $P_r\left({S_{L}},{{S_{H}}}\right)$ denotes the received signal strength.
The calculation formula is as follows: 
\begin{align}
    P_r\left({S_{L}},{{S_{H}}}\right) = {P_t G_t G_r}{\left({\frac{4\pi{Q_{(S_{L},S_{H})}}f}{c}}\right)}^{-2}
\end{align}
where $P_t$ denotes the transmit power, $G_t$ denotes the receive gain, $G_r$ denotes the transmit gain, $f$ denotes the carrier frequency. 
So the transmission time between LEO and HEO satellites is 
\begin{align}
t^{tr}_n = \frac{f^2_{{E_n},P_n}(D_n)}{L^{ISL}_{latency}}
\end{align}
Given that adjacent tasks may be generated from different satellites, there exists at most one task queue on each satellite. we denote that task $T_b$ and $T_n$ are generated adjacently on the same satellite, one after another i.e., $T_b$ followed by $T_n$. 
Further, we can compute that the time point when LEO satellite begins to process $T_n$ is $t^{S_1}_n = \max({t^{A}_n},{t^{O_1}_b})$, the time point when LEO satellite gets ready to transmit $T_n$ is $t^{O_1}_n = t^{S_1}_n + t^{P_1}_n$, the time point when HEO satellite starts to inference $T_n$ is $t^{S_2}_n = \max({t^{O_1}_n + t^{tr_1}_n},{t^{O_2}_b})$, the time point when $T_n$ is finished is $t^{O_2}_n = t^{S_2}_n + t^{P_2}_n$. Then, the time point when the LEO processes the task $T_n$ finally is $t^{S_3}_n = \max({t^{O_2}_n + t^{tr_2}_n},{t^{O_3}_b})$, the time point when $T_n$ is finished is $t^{O_3}_n = t^{S_3}_n + t^{P_3}_n$. Finally we compute the total inference time for $T_n$ is $T^P_n = t^{O_3}_n - t^{A}_n$.

\subsection{Problem Formulation}
For the multi-exit model, each EE point has its specific level of accuracy due to different computing amounts. 
Thus the accuracy of EE point $E_n$ is donated by $A_{E_n}$, where $0 \le A_{E_n} < 1$. In this case, $A_{E_n} = 0$ if and only if $E_n = 0$. Feng et al. 
designed a cost computing function that tradeoff inference performance and latency by introducing a novel weight parameter \cite{fang2018nestdnn}. 
This enables the selection of a descendant model that satisfies the specifications of resource-constrained satellites.
Similarly, striving to enhance task accuracy while completing as many tasks as feasible, we introduce an exponential function and define task gain $G_n$ of the $n$-th task as
\begin{align}
  G_n = \left\lceil{A_n}\right\rceil + \frac{\alpha}{1+{e^{{-\beta}{({A_n-A_{min}})}}}} 
\end{align}
where rounding up the inference accuracy $\left\lceil{A_n}\right\rceil$ quantifies the number of tasks completed in the system, while the \textit{Sigmoid} function associated with the inference accuracy represents the user's satisfaction with task performance.
Specifically, $A_{min}$ is the minimum branchy model inference accuracy. 
We introduce two parameters $\alpha$ \textgreater 0 and $\beta$ \textgreater 0, where the performance improvement factor $\alpha$ controls the significance of task inference performance. And $\beta$ is used to finely adjust the influence of changes in accuracy.

Finally, we formulate the task stream with $N$ tasks in the system as an optimization problem:
\begin{gather*}
    \mathop{max}\limits_{E_n,P_n} \sum_{n=1}^{M}{G_n}
\end{gather*}
\begin{align}
    s.t.~~& t^{S_1}_1 = t^A_1, \\
    &  t^{O_3}_n \leqslant t^{A}_n + L_n,  \notag \\
    &  0 \leqslant E_n \leqslant M,  \notag \\
    &  0 \leqslant P_n \leqslant M_{E_n}. \notag
\end{align}
where $n \in N$, and the first constraint initializes the first task, the second constraint offers the inference time limitation of the task, the last constraints provide the feasible range of $E_n$ and $P_n$, respectively.

\subsection{Problem Analysis}

The objective of this problem is to efficiently tackle the adaptive task stream challenge, involving the tradeoff between inference performance and task latency. This is achieved by implementing efficient partitioning and offloading techniques for branch DNN models across the entire inference process. In this case, we consider the total number of combinations of EE and partition points as $H$. When selecting the $h$th combination of these points, the total inference time and task gain of the $m$th task are $P^m_h$ and $G^m_h$, respectively.
Here we transform the formulated problem to a group knapsack problem with $H$ items, where each item need to be packed into the knapsack.
The volume and value of the $h$th item from the $m$th groups correspond to $P^m_h$ and $G^m_h$, respectively. The total capacity of the entire backpack is $t^A_M + L_M$. Therefore, the formulated problem maximizes the total value of items that can be placed into a backpack. Hence, the problem addressed in this work is classified as NP-hard.

\begin{algorithm}[t]
    \caption{Task Gain-aware Decision}
    \label{alg:offline_decision}
    \begin{algorithmic}[1]
    \REQUIRE $N$: the task number;\\
             $D_i$: the input data size of task $T_i$;\\
             $L_i$: the time constraint of task $T_i$;\\
             $t^A_n$: the arriving time of task $T_n$;
    \ENSURE  Maximum task gain $\mathcal{G}$
    \STATE Initialize $\mathcal{G}(0, t) = 0$ 
    $//$ Initialization
    \FOR{$i = 1$ to $N$}
        \IF{$t^A_{i-1} + L_{i-1} < t^A_i$}
            \STATE $\mathcal{G}(i-1, t^A_i) = \mathcal{G}(i-1, t^A_{i-1} + L_{i-1})$\\
            $//$ Update the recursion task gain value.
        \ENDIF
        \FOR{$j = t^A_i$ to $t^A_i + L_i$}
            \STATE Update $\mathcal{G}(i, j)$ according to Eq. (7)
        \ENDFOR
    \ENDFOR
    \end{algorithmic}
\end{algorithm}

\begin{algorithm}[t]
    \caption{Optimal Point Selection}
    \label{alg:offline_selection}
    \begin{algorithmic}[1]
    \REQUIRE $M$: the number of EE points;\\
             $\{M_k | k = 1, ..., M\}$: the number of layers of the EE points $k$;\\
             $D_n$: the input data size of $T_n$;\\
             $L_n$: the time limitation of $T_n$;\\
             $L^{ISL}_{latency}$: the inter-satellite transmission rate;\\
             $f^d_{i,j}$: the predictive model in the work
    \ENSURE Optimal $E_n$, $P_n$
    \FOR{$k = M$ to $1$}
        \STATE $E_n = k$
        \STATE $C_n = \min_{P_n=1,...,M_k} \{t^{O_3}_n - t^A_n\}$\\
               $//$ Divide the model EE points in sequence and select the smallest computing time.
        \IF{$C_n \leq L_n$ \textbf{and} $A_n \geq A^*_n$}
            \STATE \textbf{record} $E_n, P_n, A_n$
        \ENDIF
        \STATE \textbf{return} $E_n, P_n$ when $A_n$ is maximum
    \ENDFOR
    \STATE \textbf{return} \texttt{NULL}
    \end{algorithmic}
\end{algorithm}

\section{Task Gain-aware Inference Offloading Algorithm}
To achieve the maximum task gain, this work introduces a task gain-aware method designed to identify the optimal EE and partition points for the task. 

We begin by calculating the maximum gain of each task at each time slot, and ultimately obtain the overall maximum gain using Algorithm \ref{alg:offline_decision}.
In this case, we can forecast task stream information including the input data, arrival time, time limitation, and total task number. We define the maximum gain $\mathcal{G}(i,j)$ with the $i$th tasks and the limitation of $j$th time slots:

\begin{equation}
\begin{split}
    \mathcal{G}(i, j) &= \max_{t^A_i \leq t < j} \{ \mathcal{G}(i-1, j), \mathcal{G}(i-1, t) + \lceil S(D_i, j - t) \rceil + \\
& \alpha \frac{1}{1+e^{-\beta(S(D_i,j-t)-A^*_i)}} \}
\end{split}
\end{equation}

Meanwhile, $S(D,t)$ is used to denote the highest accuracy of EE points with the data size $D$ and latency constraint $t$.
Algorithm \ref{alg:offline_decision} first initializes the array $\mathcal{G}(0, t)$ ($1 \leq t \leq t^A_M + L_M$). Then, when the task $i$ arrives, if the sum of the arrival time of the previous round and the latency constraint is less than the current task's arrival time, we update $\mathcal{G}({i-1}, {t^A_i})$ to $\mathcal{G}({i-1}, {{t^A_{i-1}}+{L_{i-1}}})$. At the same time, for each time point $j$, we loop from $t_{i}^A$ to $t_{i}^A+{L_{i}}$ recursively, comparing the maximum gain of the previous round $\mathcal{G}({i-1}, j)$. Thus we can obtain the maximum $\mathcal{G}(i, j)$.

Algorithm \ref{alg:offline_selection} selects the optimal points with both the highest accuracy and time constraints through an exhaustive search of the EE and partition points.
When the task $T_n$ arrives, for each branch in multi-exit model, we select the partition point in sequence and calculate the total processing time. Then, we select the minimum total processing time and the corresponding partition point is the best choice for the current task and branch. Additionally, since each branch of the multi-exit model corresponds to a task accuracy $A_n$, we can obtain a series of $A_n$ for the current task and the corresponding branch. Among those, we select the maximum $A_n$. Finally, the corresponding EE and partition points are the $E_n$ and $P_n$ that the current task should be associated with.

\begin{figure}
  \centering
  \includegraphics[width=0.95\linewidth]{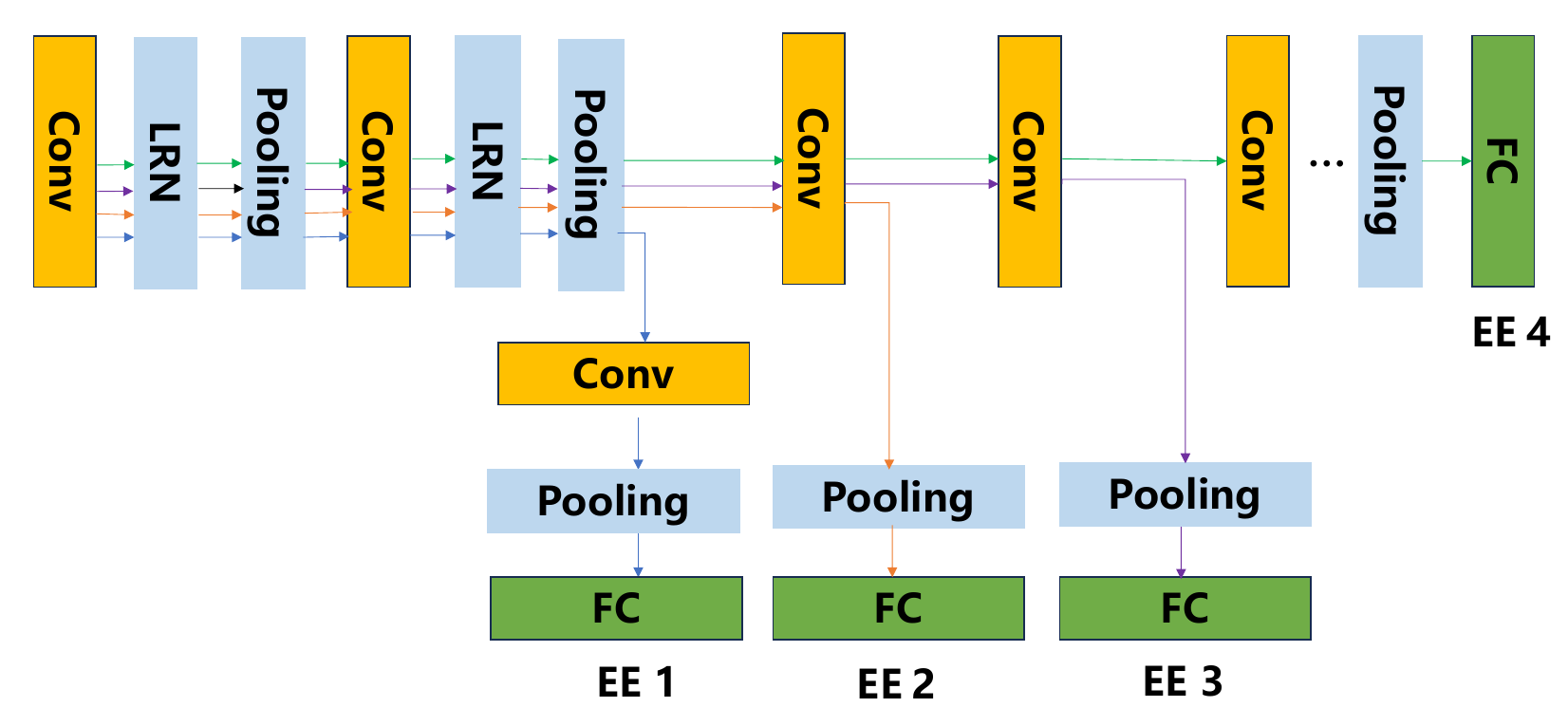}
  \caption{AlexNet model consists of 4 EE points. When we consider task failure as another particular case when starting, AlexNet model has 5 EE points.}
  \label{fig:model}
\end{figure}

\section{Simulation Evaluations}

In this section, we evaluate the performance of our proposed algorithm with the following two algorithms:
\begin{itemize}
\item Greedy: The algorithm makes the offloading decisions according to the information of the current tasks system, disregarding the interaction between adjacent tasks.
\item Random: 
The algorithm makes randomly the offloading decisions without task system information.
\end{itemize}


\newcommand{\upcite}[1]{\textsuperscript{\textsuperscript{\cite{#1}}}}
\subsection{Experiment Setup}
We use the satellite constellation StarLink and the satellite GPS system as references to simulate inter-satellite connectivity in orbit, and use various computing capabilities of LEO satellites by adjusting the number of CPUs, operating peaks, etc.
The dataset used in the simulation is CIFAR-10 and each image is categorized into one of ten classes. The dataset is also used for model training, simulating the tasks, and evaluating the proposed algorithm. Additionally, we utilize widely-used classification as our evaluated DNN-based inference application and adopt an 8-layer AlexNet model including 4 EE points. AlexNet model we consider is shown in Fig. \ref{fig:model}. When we regard task failure as another particular case when starting, this model consists of 5 EE points. Therefore, after training, the corresponding inference accuracy of 5 EE points is [0, 0.527, 0.623, 0.697, 0.743].

In the evaluation, task generation follows \textit{Bernoulli distribution} with $p =0.1$, ensuring that at most one task is generated during each time slot. The input is the images to be inferenced in the task system, where the number of images follows a normal distribution within [1,10]. Every time slot is set to 3 seconds. In addition, the hyperparameters $\alpha$ and $\beta$ are set to 0.1 and 16, respectively.

We consider three vital metrics to evaluate the proposed algorithm: the overall task gain, task completion rate, and average task latency. The overall task gain refers to Eq. (6), which considers the two task performance in the entire system. Task completion rate is calculated by dividing the number of tasks completed by the number of tasks arrived, which allows us to study the impact of system decisions on task completion from the user's perspective. The average task latency is calculated by summing the differences between the completion time of completed tasks and their arrival time, and then dividing it by the number of tasks completed, which indicates the impact of system decisions on the average task latency. 
    
\subsection{Experiment Results}

\begin{figure*}[htbp]
  \begin{minipage}[t]{0.24\textwidth}
    \centering
    \includegraphics[width=\textwidth]{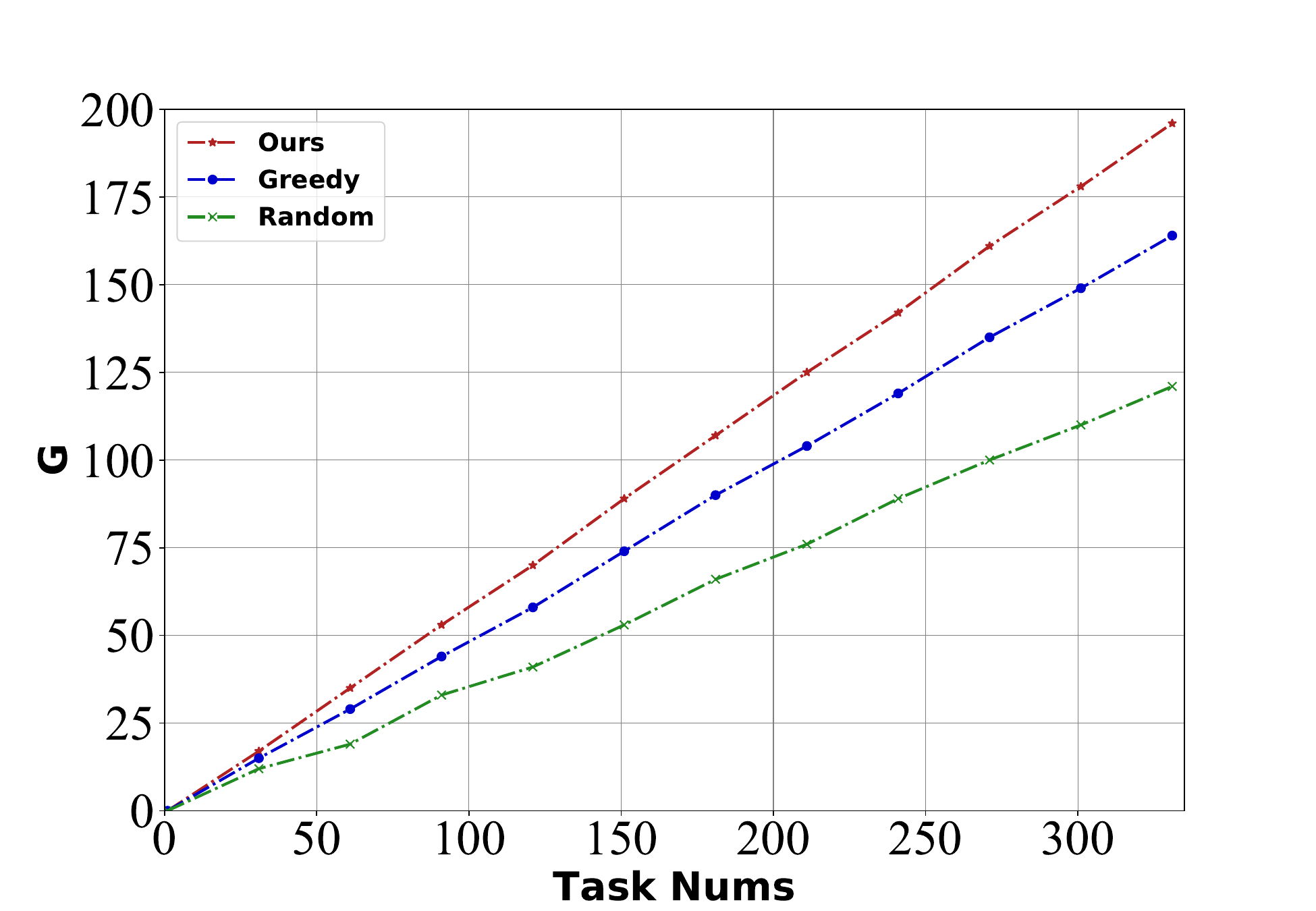}
    \caption{The total gain in different the number of tasks.}
    \label{fig:Lab2}
  \end{minipage}\hfill
  \begin{minipage}[t]{0.24\textwidth}
    \centering
    \includegraphics[width=\textwidth]{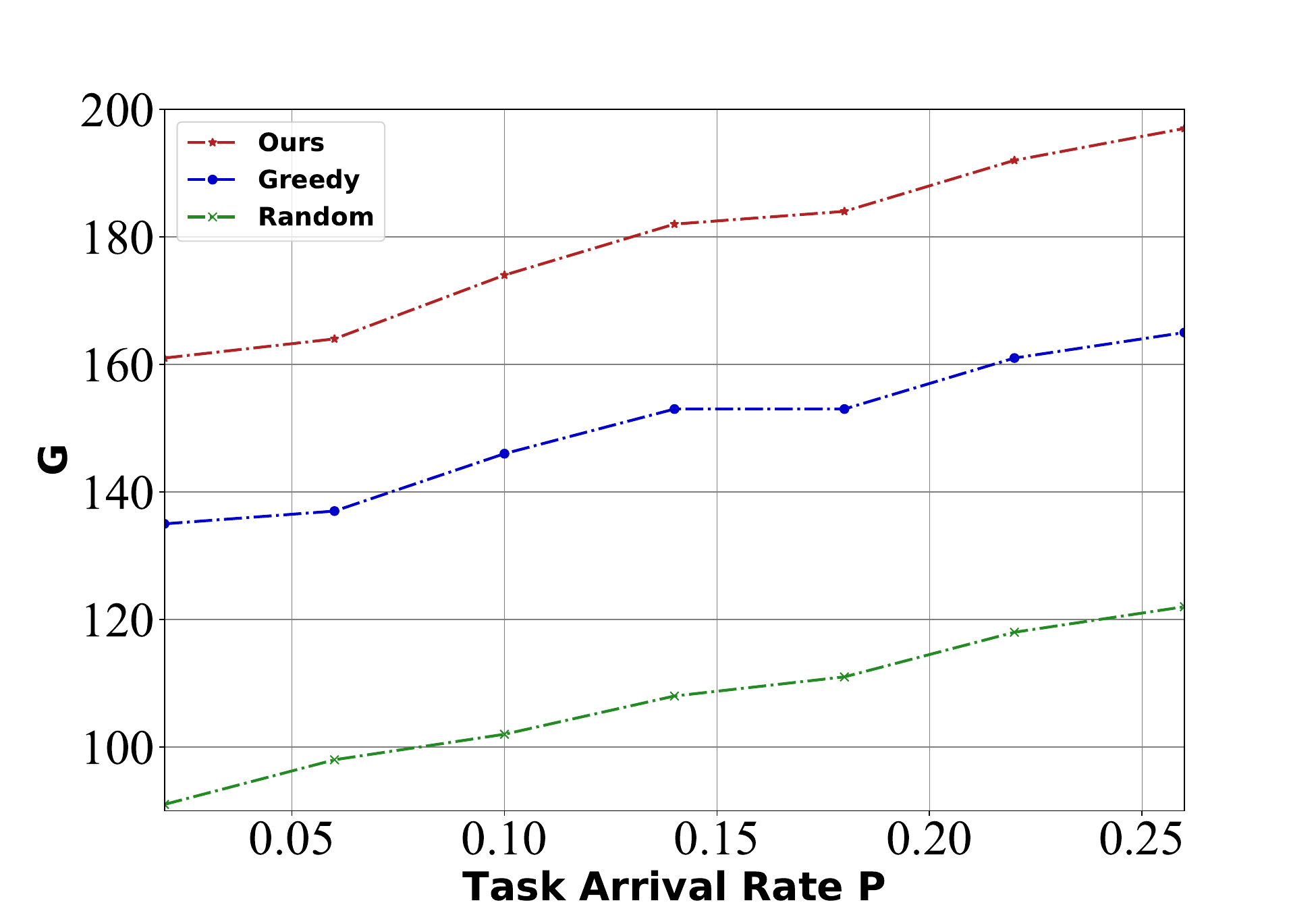}
    \caption{The total gain under the different task arrive rate $p$.}
    \label{fig:Lab1}
  \end{minipage}\hfill
  \begin{minipage}[t]{0.24\textwidth}
    \centering
    \includegraphics[width=\textwidth]{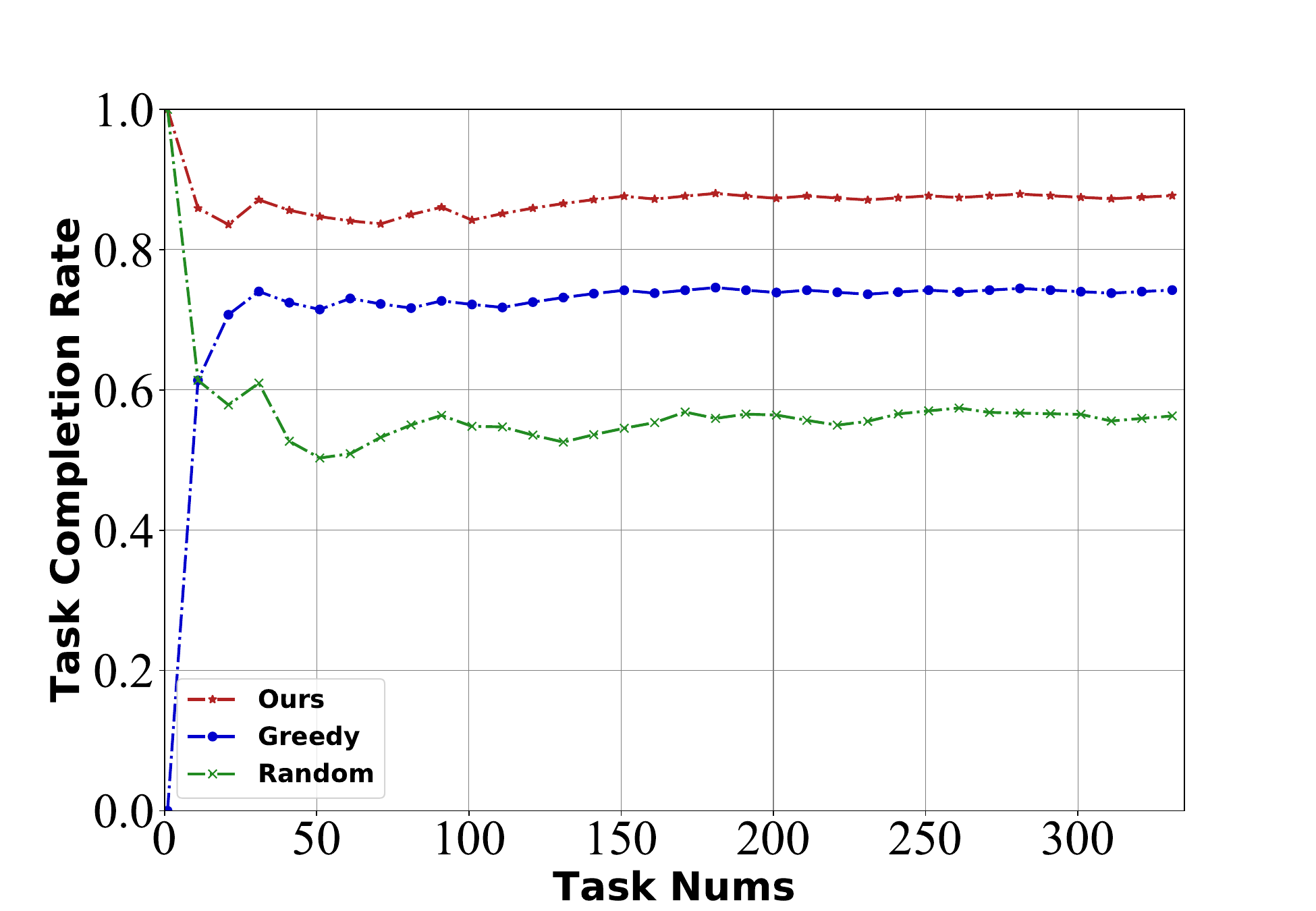}
    \caption{The task completion rate in different task numbers.}
    \label{fig:Lab3}
  \end{minipage}\hfill
  \begin{minipage}[t]{0.24\textwidth}
    \centering
    \includegraphics[width=\textwidth]{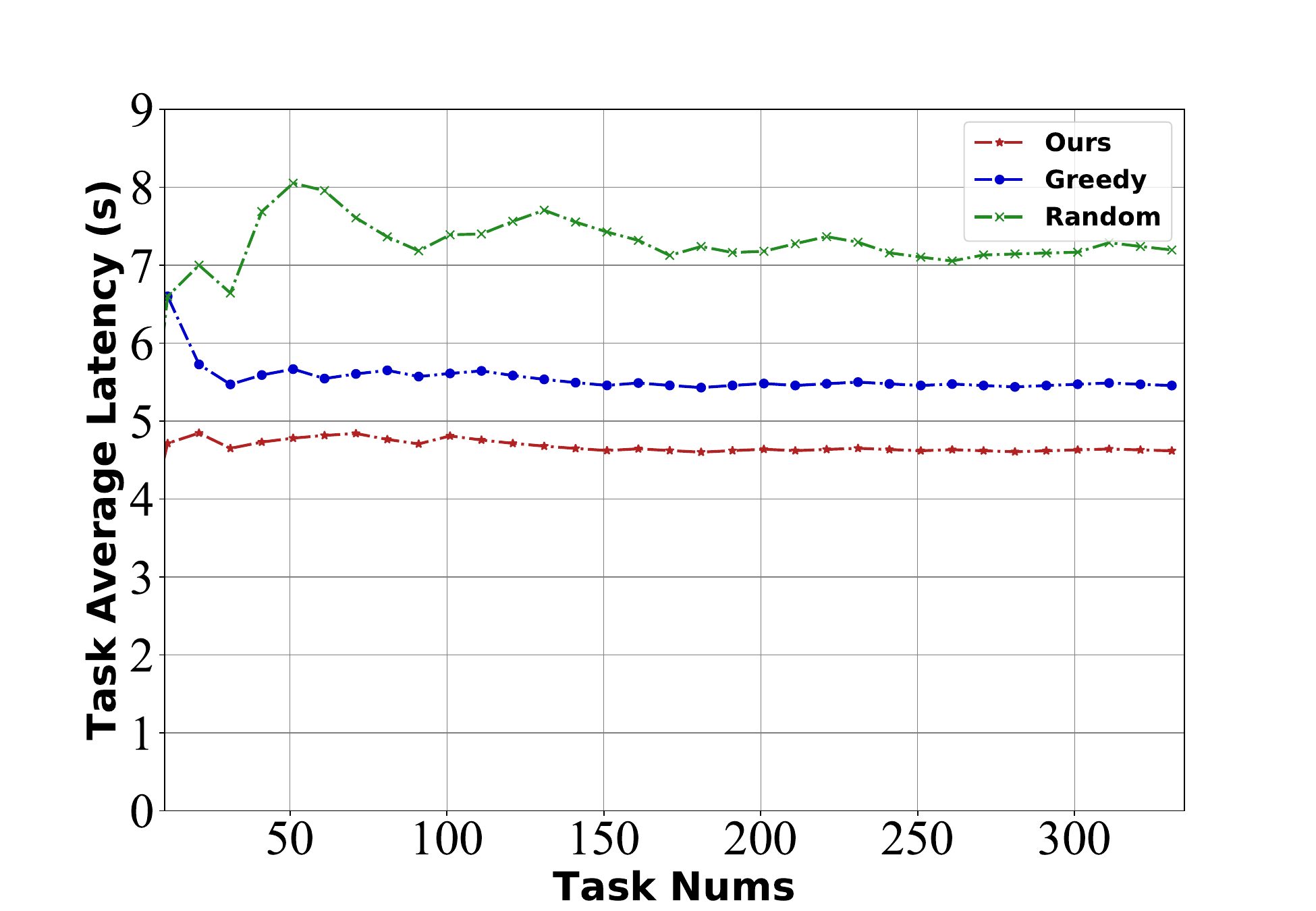}
    \caption{The task average latency in different task numbers.}
    \label{fig:Lab4}
  \end{minipage}
  \label{fig:side_by_side}
\end{figure*}

As illustrated in Fig. \ref{fig:Lab2}, our algorithm consistently demonstrates superior performance compared to the baseline algorithms. As expected, with the intensive number of task arrivals, our algorithm demonstrates approximately linear growth in the overall system objective. Notably, when the system's task arrival number is 300, our algorithm outperforms the greedy and random algorithms by 19.39\% and 61.48\%, respectively. This improvement can be attributed to our algorithm's consideration of the interdependence between adjacent tasks in the task system, which incorporates both the average accuracy of tasks and the number of completed tasks. 
Furthermore, Fig. \ref{fig:Lab1} demonstrates that our algorithm consistently outperforms the Greedy and Random algorithms, regardless of variations in the task arrival rate. Additionally, the excellence of our algorithm becomes progressively more apparent as task arrival rates increase, highlighting its remarkable performance in complex task scenarios.

Next, we investigate the task completion rate of all algorithms under varying numbers of task arrivals, as shown in Fig. \ref{fig:Lab3}, and find our algorithm's superiority over the other algorithms. When the task arrival number is 300, our algorithm achieves an 18.1\% and 55.5\% improvement compared to the Greedy and Random algorithms, respectively. This advantage originates from our algorithm's ability to determine the tradeoff between task accuracy and latency constraints. By avoiding excessive time dedicated to individual tasks, our algorithm achieves a significantly higher task completion rate.

Finally, we investigate the average task latency of all algorithms under varying numbers of task arrivals, as illustrated in Fig. \ref{fig:Lab4}. Our findings reveal that our algorithm consistently demonstrates lower latency than other algorithms. When the task arrival number is 300, our algorithm reduces latency by 35.6\% and 15.3\% compared to the greedy and random algorithms, respectively. This reduction can be attributed to our algorithm's comprehensive consideration of both average task accuracy and the number of completed tasks. 
By prioritizing efficient task completion while maintaining accuracy requirements, our algorithm effectively minimizes average task latency.

\section{CONCLUSION}

The paper proposes a framework for collaborative intelligent inference for various satellite inference tasks based on multi-exit model and model partition.
The framework aims to make an efficient trade-off between task completion and the average inference accuracy of DNN-based applications. We analyze the problem theoretically and design a task gain-aware algorithm to solve this issue. Simulation evaluations show that the proposed algorithm outperforms that of baseline algorithms, especially with intensive tasks and strict time constraints.
\bibliographystyle{IEEEtran}
\bibliography{ref}

\vspace{12pt}

\end{document}